# On the evolution of the non exchange spring behaviour to the exchange spring behaviour: A First Order Reversal Curve Analysis

Debangsu Roy*, Sreenivasulu K V and P S Anil Kumar

Department of Physics, Indian Institute of Science, Bangalore 560012, India

The magnetization behaviour of the soft Cobalt Ferrite-hard Strontium Ferrite nanocomposite is tuned from the non exchange spring nature to the exchange spring nature, by controlling the particle size of the soft Cobalt Ferrite in the Cobalt Ferrite: Strontium Ferrite (1:8) nanocomposite. The relative strength of the interaction governing the magnetization process in the nanocomposites is investigated using Henkel plot and First Order Reversal Curve (FORC) method. The FORC method has been utilized to understand the magnetization reversal behaviour as well as the extent of the irreversible magnetization present in both the nanocomposites having smaller and larger particle size of the Cobalt Ferrite. The magnetization process is primarily controlled by the domain wall movement in the nanocomposites. Using the FORC distribution in the ($H_a$, $H_b$) co-ordinate, the onset of the nucleation field, invasion of the domain wall from the soft to the hard phase, domain wall annihilation and the presence of the reversible magnetization with the applied reversal field for both the nanocomposites has been investigated. It has been found that for the composite having lower particle size of the soft phase shows a single switching behaviour corresponding to the coherent reversal of the both soft and hard phases. However, the composite having higher Cobalt Ferrite particle size shows two peak behaviour in the FORC distribution resembling individual switching of the soft and hard phases. The FORC distribution in ($H_u$, $H_c$) co-ordinate and the Henkel measurement confirms the dominant exchange interaction in the nanocomposites exhibiting exchange spring behaviour where as the occurrence of both the dipolar and exchange interaction is substantiated for the non exchange coupled nanocomposite. The asymmetric nature of the FORC distribution at $H_c$= 0 Oe for both the nanocomposites validates the intercoupled nature of the reversible and irreversible magnetizations in both the nanocomposites. It is also concluded that the contribution of the reversible magnetization is more in the nanocomposite having lower particle size of the Cobalt Ferrite compared to the nanocomposite having higher particle size of the Cobalt Ferrite.

*debangsu@physics.iisc.ernet.in

# I. Introduction:

In the recent years, increasing attention has been paid in the field of nanocomposite magnet [1,2] as it provides an integrated system comprising of components whose properties are complementary to each other. One such active field of research is the exchange spring magnet[3-7], where high saturation magnetization of the soft and the high magnetic anisotropy of the hard magnetic phases are exchange coupled in the nanometric scale. Both the experimental studies and the Micromagnetic calculations[8-11] reveal that significant improvements in terms of magnetic energy product $(BH)_{max}$ can be achieved using the exchange spring magnet. It has also been suggested that the exchange coupling between the hard and soft phase should be rigid[8] which will lead to the good squareness and higher $(BH)_{max}$. Coherent rotation of the spins at the hard-soft interface is the key for the rigid exchange coupling[9]. Thus the better understanding of the interactions present between the soft and hard phases will lead to the improvement of the nanocomposite magnet with improved magnetic properties compared to the individual soft and hard magnetic phases. The hysteresis and reversible processes are central to the investigation of the microscopic interactions[12]. In general, the competition between the reversible and the irreversible switching processes determines the magnetic interaction present in the system. In the case of exchange spring the switching field corresponding to the soft and hard phases manifests the inter-phase coupling strength. We have previously showed the effect of the soft phase ($Ni_{0.8}Zn_{0.2}Fe_2O_4$) on the coercivity mechanism[13] of the nanocomposite $Ni_{0.8}Zn_{0.2}Fe_2O_4/BaFe_{12}O_{19}$. We have also demonstrated the occurrence of the exchange spring in the case of oxide ferrites [14] and enhancement of $(BH)_{max}$ in all oxide exchange spring system[15]. However, we have not investigated the evolution of the exchange spring mechanism in these systems.

In this article, we compare the magnetization reversal processes in the hard-soft Strontium Ferrite-Cobalt Ferrite nanocomposites that exhibit exchange spring and non exchange spring nature. In order to realize the reversal mechanism, one needs to vary the strength of the magnetic interaction in the nanocomposites. In addition, one needs to have a suitable technique to quantify the effect of the interaction. The particle size of the soft phases has been varied to understand the evolution from the non exchange spring behaviour to the exchange spring behaviour in the Strontium Ferrite-Cobalt Ferrite nanocomposite. We have used the first-order reversal curve (FORC)[16-19] and Henkel Plot[20,21] technique to understand the reversal mechanism present in these two systems. The FORC method is based on the procedure described by Mayergoyz[22]. It is a versatile yet a simple technique which provides plethora of quantitative information apart from the evaluations of the interaction in the hysteretic system regardless of whether it is bulk[23], thin film[24-29], nanowire[30-32], magnetic tunnel junction[33, 34], ferroelectric switching[19, 35], patterned system[36, 37] and permanent magnets[38]. In the present work, the irreversible switching processes during the magnetization reversal have been investigated for two systems namely viz. non exchange spring and exchange spring. We have also investigated the switching field distribution (interaction field or bias field distribution) and coercive field distribution for both the systems. Using these profile distribution for both the systems, the existence of the pinning, homogeneity of the sample has been determined. The occurrence of the nucleation, domain wall annihilation processes involving the magnetization switching for both the exchange and non exchange coupled composite has been analysed with the variation of the applied reversal field. The amount of magnetization irreversibility has also been measured and it is found that the irreversibility is more when the nanocomposite is in the exchange spring regime compared to the non exchange spring regime. Furthermore the coupling between reversible and irreversible magnetization has also been investigated.

## II. Experiments:

The hard ferrite, Strontium Ferrite ($SrFe_{12}O_{19}$) has been prepared by the citrate gel method. In this method precursors of Strontium Nitrate and Ferric Nitrate molar solutions were mixed in the appropriate ratio and subsequently mixed with the citrate gel and then subjected to the heat treatment at $300^0$ C. The soft ferrite with the representative composition of $CoFe_2O_4$ has been prepared by the citrate gel method using Cobalt Nitrate and Ferric Nitrate. Firstly, the $300^0$ C heated Cobalt Ferrite was separated into two batches. One batch was heat treated at $1000\ ^0$C for 3 hours where as the other batch was kept as such. Then the $300^0$C heated Cobalt Ferrite was mixed with the $300^0$ C heated Strontium Ferrite in the weight ratio of 1:8. The obtained mixture was further heat treated at $1000^0$ C for 3 hours. This sample is named as Set A. For the second sample, the $1000^0$C heated Cobalt Ferrite was taken and mixed in the weight ratio of 1:8 with the $300^0$ C heated Strontium Ferrite. The resultant mixture was then subjected to the heat treatment at $1000^0$C for 3 hours. This sample is termed as Set B. Powder X-ray diffraction (XRD) with a Bruker D8 Advance System having Cu Kα source was used for identifying the phase and the crystal structure of the nanocomposite. Henkel measurement was performed in a Quantum Design SQUID to evaluate the quantity δM(H) in the following manner, which can be expressed as δM(H)= [$J_d(H)-J_r(\infty)+2J_r(H)$] / $J_r(\infty)$. Here, the isothermal remanent magnetization ($J_r$) curve can be measured experimentally by starting with a fresh sample and consequent application and removal of the applied field in one direction (IRM). Similarly, the dc demagnetization (DCD) remanence ($J_d$) can be measured by saturating the sample in one direction followed by subsequent application and removal of the applied field in the reverse direction. The Quantum Design PPMS was utilized for the measurements of FORC in addition to the standard major loops at the room temperature. A large number of (>10$^2$) partial hysteresis curve called First Order Reversal Curve (FORC) has been obtained using the below mentioned procedure. Initially, the sample was subjected to the positive saturation after which the field was reduced to a reversal field value of $H_a$. From this reversal field until the positive saturation, the magnetization has been measured which traces out a single First Order Reversal Curve. A suite of FORC has been measured using the mentioned procedure for a series of decreasing reversal field. It has to be noted that, equal field spacing has been maintained throughout the measurement thus filling the interior of the major loop which act as an outer boundary for the measured FORCs. The magnetization on a FORC curve at an applied field $H_b$ for a reversal field of $H_a$ is denoted by $M(H_a, H_b)$ where $H_b \geq H_a$. The FORC distribution obtained from consecutive measurements point on successive reversal curve can be defined as the mixed second order derivative[17, 39, 40] given by,

$$\rho(H_a, H_b) = -\frac{1}{2} \frac{\partial^2 M(H_a, H_b)}{\partial H_a \partial H_b} \quad (1)$$

where $H_b > H_a$. The FORC distribution and related diagram has been calculated using FORCinel which use locally-weighted regression smoothing algorithm (LOESS)[41] for the calculation. Usually it is convenient to define a new set of co-ordinates $(H_u, H_c)$ instead of $(H_a, H_b)$ [$(H_u = (H_a + H_b)/2, H_c = (H_b - H_a)/2,)$] for the representation of the FORC diagram thus rotating the FORC diagram by 45° from $(H_a, H_b)$ plane to $(H_u, H_c)$ plane[40]. We have used both the coordinate system for the discussion of the results. In this measurement both the composites were initially subjected to the maximum field of 50000 Oe which is assumed to be the ground state for both the samples, thus neglecting the effect of the metastable domain walls. Afterwards we have varied the reversal field from 2300 Oe to -7800 Oe with the field spacing of 100 Oe in the previously mentioned manner. Thus the FORC distribution can be visualized in a top down fashion and from left to right for a given reversal field. In order to quantify the difference in the interaction and reversal mechanism in the samples of Set A and Set B, a statistical analysis has been carried out for the coercivity and interaction distribution profile. The details of the findings are discussed later in this article.

**III. Results and Discussion:**

Figure 1 shows the X ray diffraction pattern for the Set A and Set B. It is clearly evident from the XRD pattern that the characteristic peaks of hard Strontium Ferrite (*) and soft Cobalt Ferrite (+) is present in both the Set A and B. No extra peak within the resolution of the XRD technique is detected in both the XRD patterns although the samples have undergone different processing prior to the heat treatment at $1000^0$C.

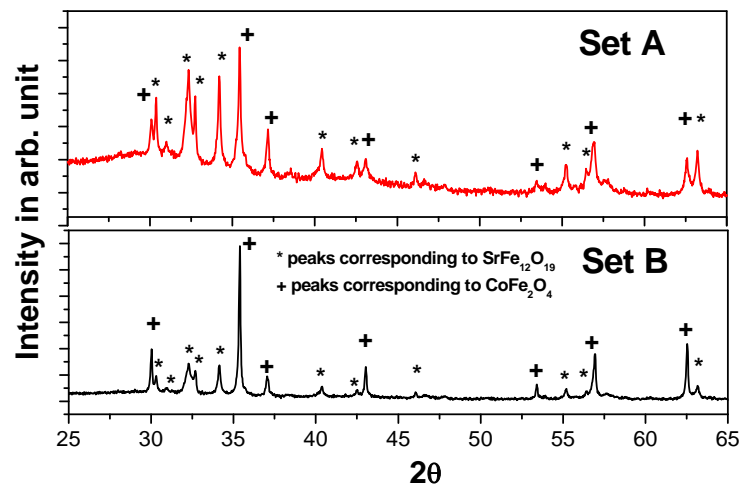

**Figure 1:** X-ray diffraction pattern for Set A and Set B. The symbol (+) and (*) represents the soft Cobalt Ferrite as well as the hard Strontium Ferrite.

From the XRD pattern of the composites, it is observed that there is no change in the peak position for the composite Set A and Set B. The broadening and the change in the intensity of the peaks between the two samples is the result of the difference in the particle size for the composites Set A and Set B. Using the Scherrer formulae[42], the average particle size corresponding to the Strontium Ferrite and Cobalt Ferrite present in Set A and Set B, has been calculated. It has been found that in both Set A and Set B, the average particle size corresponding to the hard Strontium Ferrite is >50 nm. The similarity in the average particle size for Strontium Ferrite corresponds to the fact that in both the Sets, the Strontium Ferrite particles has undergone similar processing condition. But the average particle size for the Cobalt Ferrite in both the Set A and Set B has been calculated as < 50 nm and > 70 nm respectively. This is in accordance with the fact that in Set A, as prepared Cobalt Ferrite were mixed with the as prepared Strontium Ferrite where as in Set B $1000^0C$ sintered Cobalt Ferrite were mixed with the as prepared Strontium Ferrite and subsequently heat treated at $1000^0C$ for both the mixtures. This result is corroborated with the Scanning Electron Microscopy images. This confirms the existence of the two independent major phases in the composite, without any chemical reaction.

Figure 2 shows the magnetic hysteresis loop for both the Set A and Set B. The inset shows the enlarged view of the magnetic hysteresis loop at low field.

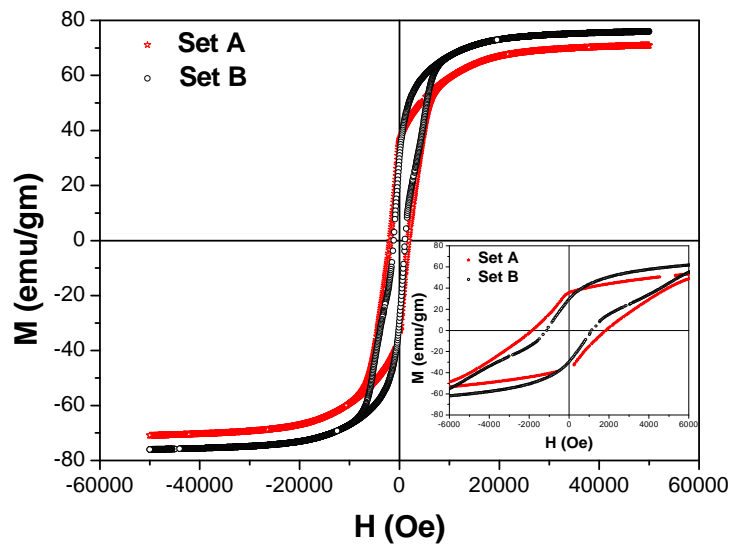

**Figure 2**: Magnetization vs. Applied Magnetic Field for the nanocomposites Set A and Set B. Inset shows the zoomed view of the magnetization loop for both the sets.

From the inset of the figure 2, it is seen that the coercivity and remanence for both Set A and Set B are 1850 Oe, 50% and 1150 Oe and 39% respectively. It is observed that in the case of the nanocomposite Set A, though crystallographically it showed two phase behaviour but magnetically it gives a good single phase behaviour. This suggests that the magnetic hard and soft phases are well exchange coupled to each other which will be demonstrated later in this article. But for Set B, the magnetic loop shows a two step hysteresis loop. This corresponds to the fact that magnetic reversal consist of two stage processes i.e. individual switching of the soft and hard phases , suggesting the absence of the exchange coupling between the hard and soft phase. In fact, the magnetic property of the two phase nanocomposite magnet depends on the size of the grains, their distribution and their shape[9]. In addition to this, the exchange as well as the dipolar interaction plays a major role for the determination of the magnetic property.

As the composite is a mixture of the soft and hard phases, three types of magnetic interaction between the soft and hard grains can be considered. The major one is the exchange interaction between the soft and hard phase where as the others two are between the soft and soft phases and between the hard and hard phases which are dipolar in nature[9]. It has to be noted that, the magnetocrystalline energy and the respective anisotropy direction of the hard and soft phases determine the extent of the exchange interaction in the isotropic nanocomposite. When an external magnetic field is applied, then it tries to align all the domains in its own direction to make it more energetically favourable. Thus there will be a competition between the magnetocrystalline energy, the exchange interaction and the interaction between the applied field and the magnetic moments of the nanocomposite. Since the soft phase is having less magnetocrystalline anisotropy energy compared to that of the hard phase, it can be easily aligned in the direction of the applied field. But in the nanocomposite, if both the phases are sufficiently exchange coupled, then the magnetic moment of the hard grains will try to prevent it. So the relative strength of the exchange interaction between the soft and hard phases acts as a tuning factor for determining the magnetic property of the nanocomposite. If one considers dipolar interaction as well, then along with the hard and soft exchange interaction, the competing dipolar interaction also decides the magnetization in the soft grains. So, the relative importance of the exchange and dipolar interaction in the composite becomes necessary for better understanding of the system and thus the same can be understood if one obtains the Henkel plot. Figure 3 shows the Henkel plot for the Set A and Set B. The positive value of the δM(H), for Set A in the figure 3 corresponds to the fact that the exchange interaction between the soft and hard phases is dominant[20] in the nanocomposite. Whereas there is a crossover from the negative to positive value of the δM(H) for Set B which suggests that until a field range ~1200 Oe, the dipolar interaction is more pronounced. This also supports the fact that the remanence in Set A is higher than the remanence in Set B. Generally in the Henkel plot, the value of the applied reversal field at which the δM(H) shows the maximum value is close to the coercive field of the system[43]. From the figure 3, the value of the reversal field for which δM(H) value becomes maximum, has been found as ~2500 Oe and 3000 Oe for Set A and Set B respectively. The deviation of the reversal field values from the respective coercive field values for Set A and Set B could be the manifestation of the magnetically mixed phase.

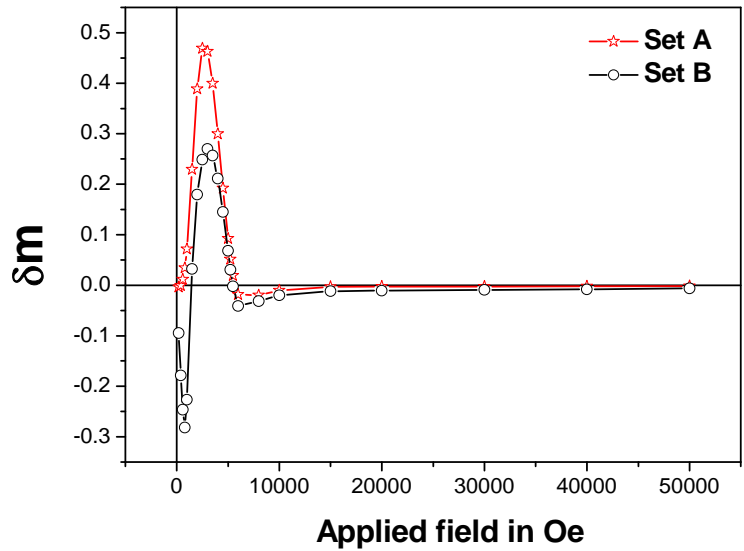

**Figure 3**: Variation of the δm vs. Applied Field for the nanocomposite Set A and Set B.

In addition there arise some limitation of Henkel plot analysis which is based on the assumption that 1. All the switching is a result of coherent rotation and 2. Domain wall does not exist in the thermally demagnetized state[44]. This assumption may not be completely valid for the systems we are investigating as the hard phase could be multi-domain in nature in both the Set A and Set B. It has also been suggested in the literature that the Henkel plot cannot differentiate between the mean interaction field and local field variance. According to Bertotti and Brasso et al [45], the variance in the local interaction field with the mean interaction field can lead to positive as well as the negative δM(H) curve depending on the relative weight of the variance and mean interaction field. Thus a positive mean interaction field does not always ensure the positive δM(H) curve if one considers the local variance as well. Thus the Henkel plot measurement is a qualitative tool for the investigation of the exchange interaction present in our system. To overcome this difficulty as well as to study the different magnetization processes occurring during the magnetization reversal, we have obtained the FORC diagram for the nanocomposite Set A and Set B.

Figure 4(a) and (c) shows the experimentally obtained first order reversal curve for the Set A and Set B where the major hysteresis loop delineates the outer boundary for the FORC curve.

It is evident from the equation (1), that FORC distribution becomes non zero when the magnetization reversal involves irreversible switching[39]. Similarly the reversible processes occurring during magnetization reversal will correspond to the zero FORC distribution[46]. In case of reversible magnetic switching, there will be no change in the magnetization value $M(H_a, H_b)$ while going from one reversal point to another considering reversible magnetization switching. Thus, the magnetization will solely depend on the applied field $H_b$, making the FORC distribution zero. The inset of the figure 4(a) and (c) shows five different point of the reversal in the major hysteresis loop which will be discussed in the FORC context. We will be discussing the respective stages of the reversal for the nanocomposite Set A and Set B separately. For Set A, the reversal is initiated by the formation of the domain wall and the successive movements of them in the soft phase[4]. This is reflected in the point 1 (line 1, around +1000 Oe ) of the major loop (inset 4(a) ). Here, we are assuming that soft and hard phases are exchange coupled through their phase boundaries. The corresponding horizontal line scan at that $H_a$ along $H_b$, representing FORC distribution ρ has been shown in the figure 4(b). The zero value of ρ corresponds to the fact that magnetization change is reversible in nature. This has also been observed by the closeness and overlap of the successive reversal curves. After the start of the domain wall movement with decreasing reversal field, the FORC distribution becoming non zero near to the zero reverse field indicating the onset of the irreversible processes. The irreversible process peaks around a reversal field value of -370 Oe (point 2 in the inset of the figure 4(a) and line 2 at figure 4(b)) and the same can be observed in the major loop with a decrease in the magnetization. This irreversible process can also be visualized from the uneven separation of the successive FORCs[27]. This is happening as the domains present in the soft phase have suddenly started to invade the hard phases leading to the irreversible switching of both the hard and soft phases. This corresponds to the fact that both the soft and hard phases are exchange coupled with each other. This field corresponds to the nucleation field of the composite. This is similar to the switching behaviour of FeNi/FePt exchange spring bilayer as described by Davies et al[26]. The FORC distribution is apparently without any new feature apart from existing non zero distribution between lines 2 and 3 (-370 Oe > $H_a$ >-1800 Oe ) as evident in the figure 4(b). It has been found that between line 3 and line 4(point around -3700 Oe) in the figure 4(b), the FORC distribution shows number of negative peaks in a positive background. In this reversal field range, the slope of the reversal curve initially stay constant but around $H_b$>0 Oe shows an abrupt increase thus resulting in the positive background-negative peak in the FORC distribution[27, 47].

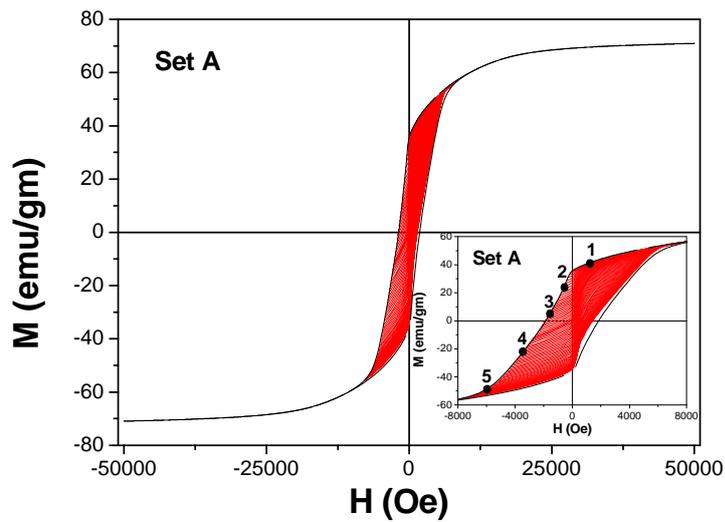

**Figure 4(a):** Magnetization loop for the nanocomposite Set A delineating FORC curves. Inset shows the zoomed view of the magnetization loop for the Set A showing 5 different reversal points.

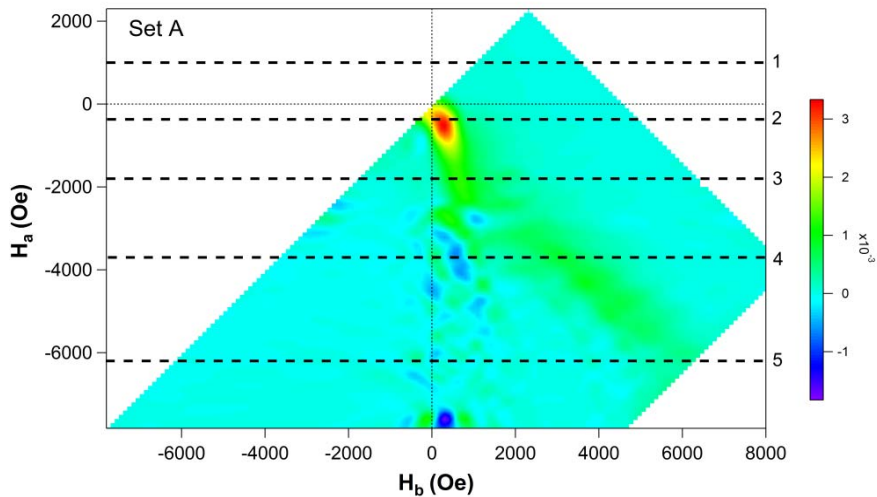

**Figure 4(b):** FORC distributions for the nanocomposite Set A in the $H_a$-$H_b$ co-ordinate.

The occurrence of the negative-positive pairing is a result of the decrease and increase of the reversal field susceptibility as the domain state responds differently with the applied field. This is the onset of the domain annihilation as the nanocomposite approaches negative saturation. Until the point -6200 Oe, the FORC distribution shows the presence of the irreversible magnetization. If one considers that the nanocomposite has reached its negative saturation then the irreversible switching will be over and the FORC distribution will be zero and nearly overlap the region Ha> -370 Oe. Since the application of the reversal field is constrained by the experimental limitation (lack of higher field to saturate the magnetization) we could only achieve the state where the reversal field corresponds to the approach to the negative saturation of the composite Set A. So for the field range Ha< -6200 Oe, we could observe a reversible magnetization change as well as discrete positive negative pair of the FORC distribution.

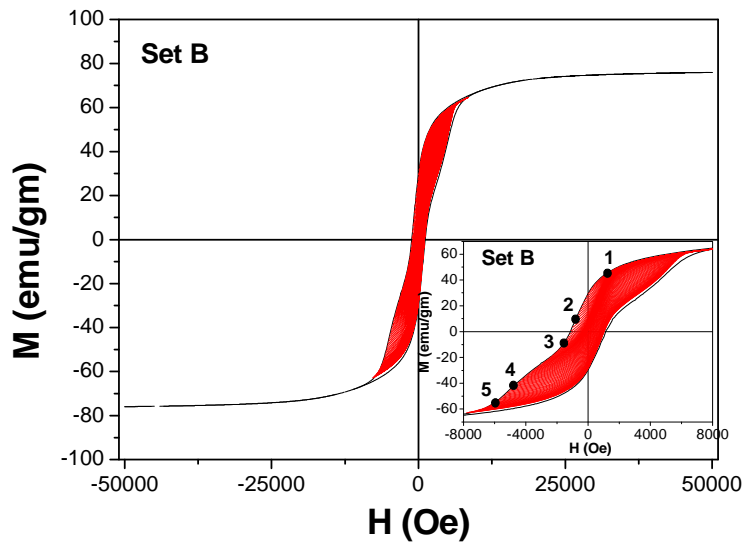

**Figure 4(c):** Magnetization loop for the nanocomposite Set B delineating FORC curves. Inset shows the zoomed view of the magnetization loop for the Set A showing 5 different reversal points.

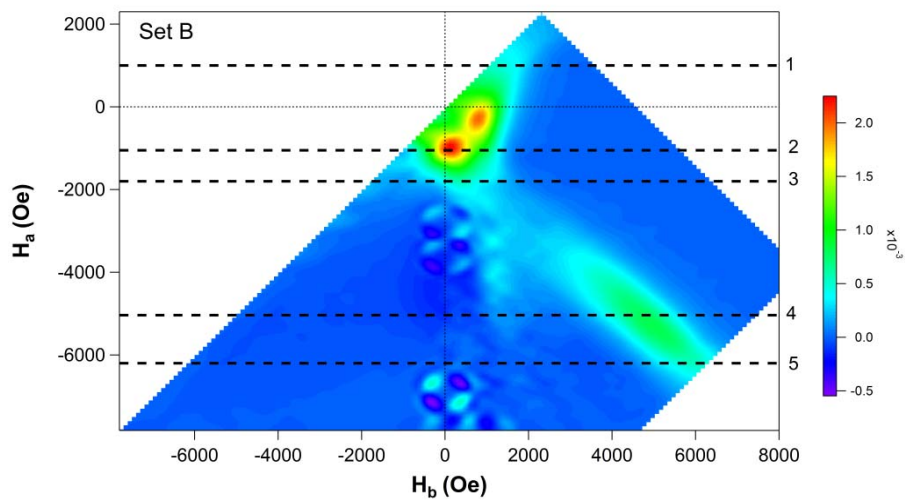

**Figure 4(d):** FORC distributions for the nanocomposite Set B in the $H_a$-$H_b$ co-ordinate.

Figure 4(d) shows the FORC distribution of Set B in ($H_a$, $H_b$) co-ordinate. In Set B, prior to the Line 1 (point +1000 Oe in the inset of the figure 4(c) ) in the figure 4(d), a very low value of FORC distribution corresponds to the reversible switching of the moment is seen. The onset of irreversible switching is demonstrated around the line 1. ($H_a$ < +1000 Oe). This corresponds to the individual switching of the soft phase present in the system. This irreversible switching shoots up between the line 1 and line 2 (point 2, -1050 Oe), in the FORC distribution as evident from the figure 4(d). The first peak corresponding to the irreversible switching in the nanocomposite Set B is a result of the individual switching of the soft phase (around ~-300 Oe) and correspondingly the second peak is because of the switching of hard phase (point 2). This irreversibility is also evident from the fact that the gap of FORC curve widens up during this reverse field range. The line scan around -300 Oe and -1050 Oe reveals that, the strength of irreversibility is more in case of -1050 Oe reversal field compared to -300 Oe. This corresponds to the fact that both the soft and hard phase are not sufficiently exchange coupled in the nanocomposite Set B and does individual switching as compared to Set A which gives rise to a single peak of irreversibility revealing coherent rotation of the soft and hard phases. Generally domain wall movement require less energy compared to the coherent rotation of the magnetization. So as the field is reversed after saturation without changing the sign of the applied field, the domain wall starts moving from one energy minima to the other. In this process if the system reaches nucleation and consequently the applied field is ramped back, the system will not achieve the initial condition thus giving rise to an irreversible magnetization. Thus, it can be concluded that the domain nucleation is happening at the field range of < +1000 Oe. Thus with the decreasing reversal field, the composite mostly composed of "down" domains compared to the "up" domains thus resulting an abrupt change in magnetization around 0 Oe. This irreversible process of converting "up" domains into "down" domain configuration continues until a reversal field value of the ~ -1800 Oe as evident from the FORC distribution shown in figure 4(d). (between line 2 and 3). This also results in the two stage hysteresis loop. Between line 3 and line 4 (point -5050 Oe), the FORC distribution shows a negligible irreversible processes but a non zero tails of the FORC distribution arises when the reversible field corresponds to $H_a$ > -5050 Oe. The reappearance of the significant irreversible processes is because of the onset of the annihilation of the domains which continues until negative saturation. This feature is evident until the line 5 (point 5 in the Major loop, $H_a$ < -6200 Oe.). Afterwards the nanocomposite shows a reversible behaviour for the rest of the reversal field value until -7800 Oe. So, it has been found that in case of Set B, two peaks corresponding to the irreversible change in magnetization occurs compared to the single peak of the irreversibility observed in Set A.

In order to discuss the distribution of the magnetic characteristics like interaction mechanism, coercivity distribution in the nanocomposite of Set A and Set B, we have obtained the FORC distribution as a contour plot in ($H_u$, $H_c$) coordinate. Figure 5(a) and 6(a) shows the FORC distribution for Set A and Set B in ($H_u$, $H_c$) space with the corresponding colour scale kept at the side by side as a measure of the FORC distribution.

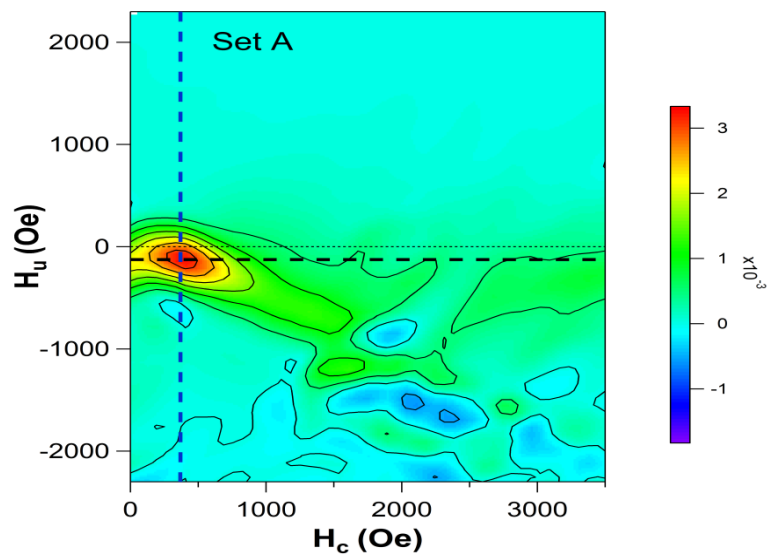

**Figure 5(a):** FORC distributions for the nanocomposite Set A in the $H_u$- $H_c$ coordinate.

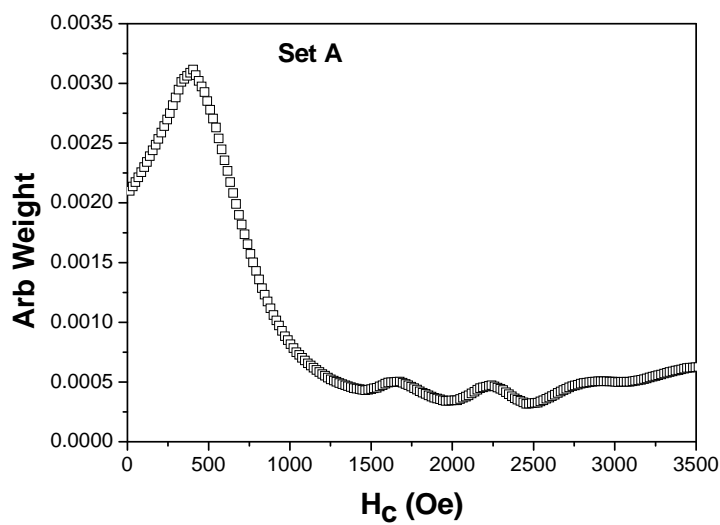

**Figure 5(b):** Coercivity profile for Set A.

In both the cases, the maximum weight of the ρ is indicated by the red colour where as the minimum is shown in the blue colour. Generally, the projection of ρ onto the $H_c$ axes, can be characterized as coercivity distribution profile. This in turn depends on the respective anisotropy distribution, size variation, defect as well as the homogeneity in the nanocomposite[23].

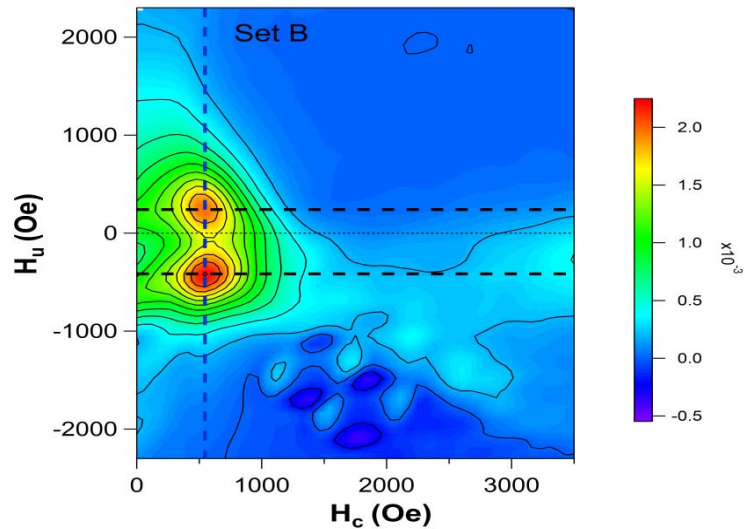

**Figure 6(a):** FORC distributions for the nanocomposite Set B in the $H_u$- $H_c$ coordinate.

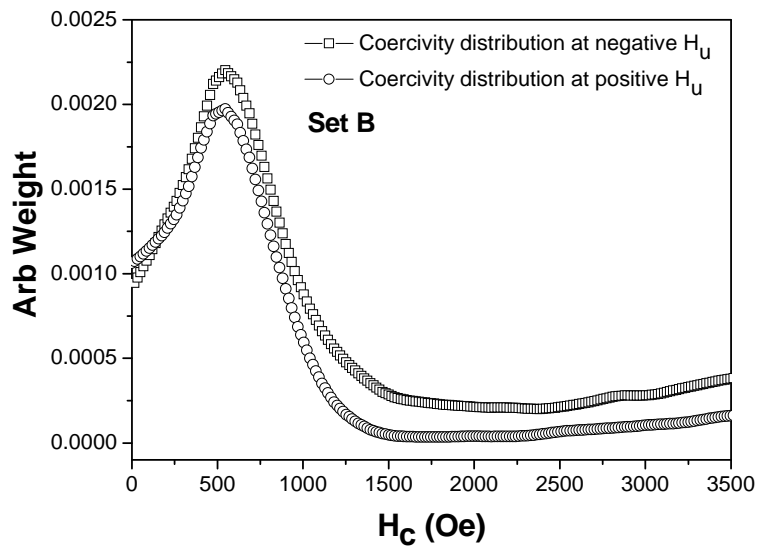

**Figure 6(b):** Coercivity profile for Set B.

Figure 5(b) and 6(b) correspond to the coercive field distribution profile for Set A and Set B which was obtained at the field values indicated by the black lines through the maxima of the

FORC distribution. Similarly the projection of ρ on to $H_u$ axis at a particular coercive field value can be visualized as the distribution of the interaction field strength. In the context of the investigated nanocomposite of Set A and Set B, this interaction field profile is a characteristic of the strength of the exchange coupling between soft and hard phases as well as the dipolar interaction present between soft-soft and hard- hard phases[23]. The figure 7 shows the respective variation of the interaction field profile for Set A and Set B.

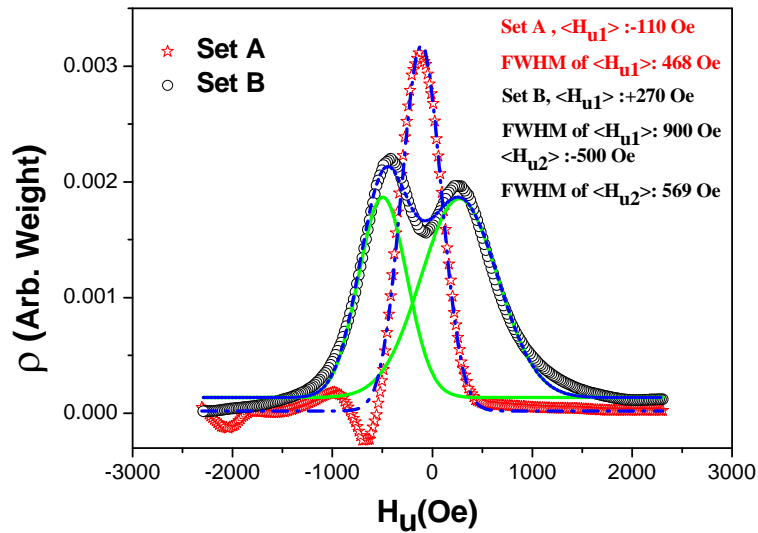

**Figure 7:** Interaction field profile for Set A and Set B. The blue line corresponds to the Gaussian fitting in both the cases. The fitting parameters are describes inside the graph.

It has been found from the FORC distribution for Set A and Set B in figure 5(a) and 6(a), that the contour diverges from the $H_c$= 0 axis and are much more pronounced at the lower field value of $H_c$. This kind of contour patterning suggests that the investigated nanocomposite is multi-domain in nature[40]. According to Robert et al[40], the origin for the diverging contour pattern can be related to the domain wall pinning and domain wall nucleation and annihilation. For the nanocomposite Set A, the maximum value of $\rho$ is 3.1 X $10^{-3}$ where as for Set B it is 2.2 X $10^{-3}$. Generally the small concentration of the ferromagnetic phases present in the system will lead to the smaller value of $\rho$[46]. Thus the higher value of the $\rho$ in Set A compared to Set B indicates the presence of lesser amount of ferromagnetic interaction in Set B compared to Set A thus giving rises to lesser irreversibility. To understand the variance of coercive and interaction field distribution in Set A and Set B, a statistical analysis has been performed[48]. The bias or interaction field distribution which was obtained at the field values indicated by the blue lines through the maxima of the FORC distribution in figure 5(a) and 6(a) for Set A and Set B has been fitted considering Gaussian distribution. The fitting has been shown in the figure 7. It has to be noted that for Set B since the distribution of ρ shows two peaks, we have considered double peak Gaussian for fitting the Set B distribution. It has been found from the figure 5(b) and 6(b) respectively, that the average coercivity for Set B is ~527 Oe while the Set A shows average coercive field as ~378 Oe. From the figure 5(a), It has also been found that the maximum FORC distribution in case of Set A is centred around $H_u$= -110 Oe and is tilted towards negative $H_u$ where as for Set B the variation of ρ shows two maxima as evident in the figure 6(a). The first maxima lies around $H_u$ = 270 Oe while the second maxima is concentrated at $H_u$ = -500 Oe. The pattern for Set A is in well agreement with the FORC distribution as previously showed for polycrystalline single phase LSCO with doping x=0.30[23] indicating the presence of the long range ferromagnetic interaction, reduction in pinning and multi-domain type reversal. According to C R Pike et al[17], theoretically when an mean interacting field was introduced in the assembly of model non interacting single domain particles, the peak of the FORC distribution generally displaces off the $H_u$=0 axis depending on the nature of the mean field interaction. If the interaction is dipolar in nature, then the distribution goes upward off the $H_u$= 0 axis thus giving a maxima centred at positive $H_u$. But the exchange interaction causes a negative shift of the FORC distribution which causes the FORC distribution peaks at negative $H_u$. So we can conclude that in Set A the mean interacting field is exchange in nature but in Set B both dipolar as well as the exchange type of interaction is present. This is in well agreement with the Henkel plot showed in figure 4. However we have also found that, negative shift of the FORC distribution is more pronounced in Set B compared to Set A. This will be explained in conjunction with the reversible magnetization present in the system. It has also been found from the figure 7, that the spread in the interaction field distribution for Set A is less compared to the spread in both the peaks for Set B. Generally the extent of the spread can be used as a measure of the extent of the domain wall pinning, nucleation and annihilation present[40] in the system. Thus for Set A, the domain wall pinning is less pronounced compared to Set B. This corroborates well with the higher average coercive field value for Set B compared to Set A. The origin of these pinning centres could be the grain boundary[13] present between soft and hard phases. The extent of the FORC distribution ρ can be related to the ferromagnetic component (strength of the interaction) present in the system. Thus one can consider a lesser magnitude of the $\rho$ for a system which contains weak ferromagnetic component. Generally by integrating $\rho(H_u, H_c)$ over the $(H_u, H_c)$ coordinate, one can visualize the part of the system which has taken part in the irreversible switching. Thus irreversible magnetization $M_{irr}$ can be formulated as $M_{irr} = \int \rho(H_u, H_c) dH_u dH_c \approx \sum \rho(H_u, H_c) \Delta H_u \Delta H_c$. Here $\Delta H_u, \Delta H_c$ corresponds to the relative field spacing of the $H_u$, $H_c$ coordinate. We have obtained the fraction of the irreversible magnetization ($M_{irr}/M_s$) after proper scaling of the reversal curve data for Set A and Set B for significant comparison. The fraction of the irreversible magnetization for Set A and Set B has been obtained as 69% and 55% respectively. The greater value of the quantity ($M_{irr}/M_s$) for Set A compared to the Set B can be correlated to the dominant exchange coupling between the soft and hard phase in Set A in comparison to the nanocomposite Set B. This information corroborate well with the Henkel plot depicted in the figure 4 where for Set A the magnitude of the δM(H) lies above Set B. This quantitative finding is also in agreement with the previous depiction of FORC distribution in ($H_a$, $H_b$) space. It has to be noted that the calculated $M_{irr}$ cannot be accounted for the entire saturation magnetization of the nanocomposite Set A and Set B. This deviation is because of the fact that we did not consider the contribution from the reversible magnetization during our calculation[39, 47].

In order to understand the reversible magnetization contribution during the magnetization reversal process, we have obtained the FORC distribution for both Set A and Set B at $H_c$= 0 Oe which have contribution from the pure reversal process present in the system. This distribution is termed as "reversible ridge"[39] and is shown in the figure 8.

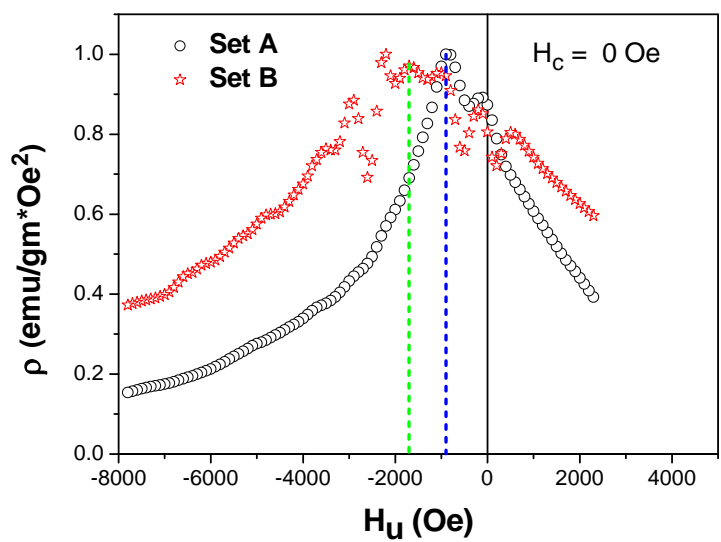

**Figure 8:** Reversible ridge for the nanocomposites Set A and Set B.

It has been found that for both the Set A and Set B, the reversible ridge as a function of $H_u$ is not symmetric about $H_u$. According to C R Pike[39], if the reversible magnetization is not coupled to the irreversible part, then the reversible ridge should be symmetric about $H_u$. Thus the asymmetry of the reversible ridge for Set A and Set B corresponds to the fact that both the irreversible and reversible magnetization is coupled with each other. This can be explained by considering the curvilinear hysteron models[39] as described in literature. According to this theoretical prediction, the upper and lower branch susceptibility of the curvilinear hysteron is different from each other thus resulting in an asymmetry of reversible ridge about $H_u$ axis. Figure 8 shows that the peak of the "reversible ridge" for Set A and Set B lay at -1700 Oe and -900 Oe respectively. This feature is indeed interesting as the interaction field for the reversible magnetization is greater in case for Set A than in Set B. Thus if we consider the coercivity in a system as a combination of both the interaction field resulting from reversible as well as irreversible magnetization processes without the presence of the domain wall pinning, in-homogeneity, then the calculated coercivity for Set A and Set B lies in the range 1810 Oe(1700+110) and 1400 Oe (500+900). It has been found that the calculated coercivity 1810 Oe for Set A agrees well with coercivity value of 1850 Oe obtained from the major hysteresis loop. But for Set B, the calculated coercivity 1400 Oe is more compared to the obtained coercive field of 1150 Oe. This indicates the presence of domain wall pinning due to the grain boundary and inhomogeneity in Set B compared to Set A. This is in well agreement with the interaction and coercive field profile distribution obtained from reversal curves. A careful examination of the reversible ridge in figure 8 reveals that the ridge shows two maxima for Set B thus correlating two stage reversals observed in Set B form the major hysteresis loop measurement. So the contribution of the reversible magnetization towards the magnetization switching process in Set A is more pronounced than in Set B thus indicating occurrence of the exchange spring behaviour in Set A relative to the non exchange spring behaviour in Set B.

## IV. Conclusion:

In conclusion, we have successfully tailored the magnetization behaviour of the Cobalt Ferrite-Strontium Ferrite nanocomposite from non exchange spring behaviour to exchange spring, by tuning the size of the soft Cobalt Ferrite. We have achieved single magnetic hysteresis loop behaviour for Set A thus confirming the exchange spring behaviour whereas the double step hysteresis behaviour for Set B corresponds to the non exchange spring behaviour. The relative strength of the interaction governing the magnetization process in the composite Set A and Set B has been studied using Henkel plot and First Order Reversal Curve method. The FORC method has been utilized to understand the magnetization reversal behaviour as well as the extent of the irreversible magnetization present in both Set A and Set B. It has been concluded that the magnetization process is primarily controlled by the domain wall movement in the composites. Using the FORC distribution in the ($H_a$, $H_b$) co-ordinate, we could trace out the onset of the nucleation field, invasion of the domain wall from the soft to the hard phase, domain wall annihilation and the presence of the reversible magnetization with the applied reversal field for both the composite Set A and Set B. We have obtained single FORC distribution maxima for Set A at a negative interaction field axis where as Set B gives rise to two consecutive maxima, one at positive and another at negative interaction field. This result is consistent with the findings from the obtained Henkel measurement for Set A and Set B. The projection of ρ on $H_u$ and $H_c$ axis clearly shows the presence of higher domain wall pinning because of the grain boundary in Set B compared to the Set A. By projecting the FORC distribution over $H_u$ and $H_c$ axis we quantitatively calculated the irreversible magnetization taking part in the magnetization reversal. The pronounced irreversible magnetization in Set A compared to Set B indicates that the ferromagnetic interaction is more prominent in Set A relative to Set B. By analysing the FORC distribution at $H_c$=0 Oe which is asymmetric about $H_u$ for both Set A and Set B, we could confirm that reversible and irreversible magnetizations are coupled with each other in both the composites. By taking into account of the interaction field corresponding to the maxima of ρ distribution in reversible as well as irreversible process for Set A and Set B, we could calculate back the coercivity obtained from major hysteresis loop. This indicates that reversible magnetization should also be taken into account while discussing the magnetization reversal process. It was concluded that the extent of the reversible magnetization is more in Set A compared to Set B. Furthermore this result confirms that FORC method is an effective tool for the investigation of the interaction and magnetization reversal processes in the exchange spring magnet.

List of Figures:

**Figure 1:** X-ray diffraction pattern for Set A and Set B. The symbol (+) and (*) represents the soft Cobalt Ferrite as well as the hard Strontium Ferrite.

**Figure 2**: Magnetization vs. Applied Magnetic Field for the nanocomposites Set A and Set B. Inset shows the zoomed view of the magnetization loop for both the sets.

**Figure 3**: Variation of the δm vs. Applied Field for the nanocomposite Set A and Set B.

**Figure 4(a):** Magnetization loop for the nanocomposite Set A delineating FORC curves. Inset shows the zoomed view of the magnetization loop for the Set A showing 5 different reversal points.

**Figure 4(b):** FORC distributions for the nanocomposite Set A in the $H_a$-$H_b$ co-ordinate.

**Figure 4(c):** Magnetization loop for the nanocomposite Set B delineating FORC curves. Inset shows the zoomed view of the magnetization loop for the Set A showing 5 different reversal points.

**Figure 4(d):** FORC distributions for the nanocomposite Set B in the $H_a$-$H_b$ co-ordinate.

**Figure 5(a):** FORC distributions for the nanocomposite Set A in the $H_u$- $H_c$ coordinate.

**Figure 5(b):** Coercivity profile for Set A.

**Figure 6(a):** FORC distributions for the nanocomposite Set B in the $H_u$- $H_c$ coordinate.

**Figure 6(b):** Coercivity profile for Set B.

**Figure 7:** Interaction field profile for Set A and Set B. The blue line corresponds to the Gaussian fitting in both the cases. The fitting parameters are describes inside the graph.

**Figure 8:** Reversible ridge for the nanocomposites Set A and Set B.